\documentclass[sigconf]{acmart}

\usepackage{natbib}
\usepackage{graphicx}
\usepackage{multirow}
\usepackage{verbatim}
\usepackage{algorithmic}
\usepackage{amsfonts}
\usepackage{subfigure}
\usepackage{enumitem}

\usepackage{amsmath}
\usepackage[psamsfonts]{amssymb}
\usepackage{amsxtra}
\usepackage{threeparttable}
\usepackage{pgfplots}

\usepackage{pifont}
\usepackage{amssymb}
\usepackage{titlesec}

\titlespacing\section{0pt}{7pt minus 2pt}{0 pt plus 2pt}
\titlespacing\subsection{0pt}{7pt minus 2pt}{0pt plus 2pt} 
\titlespacing\subsubsection{0pt}{7pt minus 2pt}{0pt plus 2pt}

\AtBeginDocument{%
  \providecommand\BibTeX{{%
    \normalfont B\kern-0.5em{\scshape i\kern-0.25em b}\kern-0.8em\TeX}}}

\setcopyright{acmcopyright}
\copyrightyear{2020}
\acmYear{2020}
\setcopyright{acmcopyright}\acmConference[MM '20]{Proceedings of the 28th ACM International Conference on Multimedia}{October 12--16, 2020}{Seattle, WA, USA}
\acmBooktitle{Proceedings of the 28th ACM International Conference on Multimedia (MM '20), October 12--16, 2020, Seattle, WA, USA}
\acmPrice{15.00}
\acmDOI{10.1145/3394171.3413579}
\acmISBN{978-1-4503-7988-5/20/10}

\begin{document}

\fancyhead{}
\title{Semi-supervised Multi-modal Emotion Recognition with Cross-Modal Distribution Matching}

\author{Jingjun Liang}
\affiliation{\institution{School of Information\\ Renmin University of China}}
\email{liangjingjun@ruc.edu.cn}

\author{Ruichen Li}
\affiliation{\institution{School of Information\\ Renmin University of China}}
\email{ruichen@ruc.edu.cn}

\author{Qin Jin}
\authornote{Corresponding Author}
\affiliation{\institution{School of Information\\ Renmin University of China}}
\email{qjin@ruc.edu.cn}


\begin{abstract}
 Automatic emotion recognition is an active research topic with wide range of applications. Due to the high manual annotation cost and inevitable label ambiguity, the development of emotion recognition dataset is limited in both scale and quality. Therefore, one of the key challenges is how to build effective models with limited data resource. Previous works have explored different approaches to tackle this challenge including data enhancement, transfer learning, and semi-supervised learning etc. However, the weakness of these existing approaches includes such as training instability, large performance loss during transfer, or marginal improvement. 
 In this work, we propose a novel  semi-supervised  multi-modal emotion recognition model based on cross-modality distribution matching, which leverages abundant unlabeled data to enhance the model training under the assumption that the inner emotional status is consistent at the utterance level across modalities.
 We conduct extensive experiments to evaluate the proposed model on two benchmark datasets, IEMOCAP and MELD. The experiment results prove that the proposed semi-supervised learning model can effectively utilize unlabeled data and combine multi-modalities to boost the emotion recognition performance, which outperforms other state-of-the-art approaches under the same condition. The proposed model also achieves competitive capacity compared with existing approaches which take advantage of additional auxiliary information such as speaker and interaction context.  
 
\end{abstract}


\maketitle

\vspace{-10pt}
\section{Introduction}

Emotion is an important part of daily interpersonal human interactions. 
Automatic recognition or detection  of human emotions have attracted much research interest in the field of computer vision, speech processing, and multimedia computing. 
Emotion recognition technology has a wide range of applications including assisting mental health analysis \cite{cesbaseline2018}, improving natural human machine interaction \cite{Fragopanagos2002Emotion}, enabling emotional robot design and intelligent education tutoring \cite{meyer2002discovering,zhang2018role} etc.

Emotion recognition can be generally categorized into two types of tasks, namely discrete (categorical) emotion recognition and continuous (dimensional) emotion recognition. The discrete emotion recognition normally divides the emotion space into several basic emotion classes such as happiness, sadness, anger and neutral etc \cite{ekman1992argument}, while the continuous emotion recognition treats emotional state as distribution in a continuous space, which is normally described by two or three dimensions such as arousal, valence and dominance \cite{russell1977evidence}. 
Although continuous emotion representation can model more flexible and complicated emotional state, it is not as easy to understand as discrete emotion representation, which is reflected in the quite high variance in continuous emotion human annotations from different annotators \cite{Scherer2013Investigating, Chang2017Learning}.
We therefore focus on the discrete emotion modeling in this work. 

\begin{figure}[t] 
\centering
\includegraphics[width=.4\textwidth]{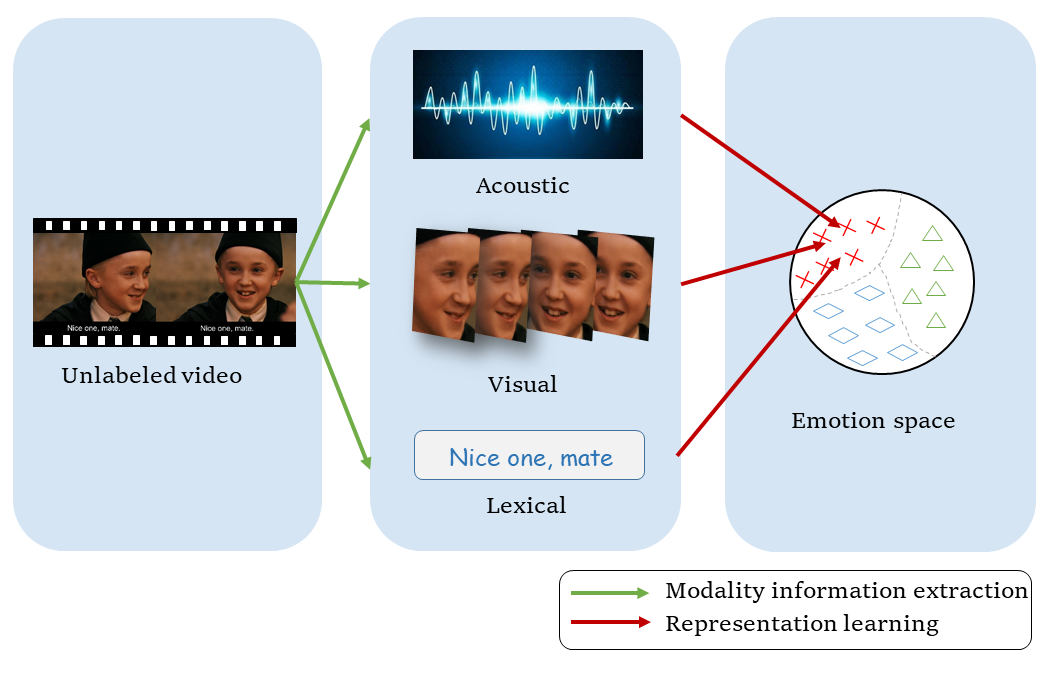}
\caption{The latent representation of acoustic, visual and lexical modalities of the same video are expected to be close when they are mapped into a common emotion space.} 
\label{concept} 
\end{figure}

We humans convey emotions in various ways including both spoken words and nonverbal behaviors, such as facial expression and body language etc \cite{gibson1993tools}. Such rich information from multi-modalities could be used to understand the emotional state \cite{manning2014stanford}. Previous research works have shown that different modalities are complementary for emotion recognition \cite{Emotion, Huang2015An}. 
Different modalities all carry emotion relevant information and how to effectively combine multiple modalities has been an active research focus.

Besides multi-modality, another challenge for emotion recognition is the limitation of supervised data.
Although we can easily collect large amount of emotional data from online social media, the emotion annotation requires heavy manual efforts and usually involves inevitable label ambiguity. Therefore shortage of high quality supervised data has been a big obstacle for developing generalized and robust emotion models.
There have been some endeavors to tackle the data shortage challenge. For example, Albanie et al. \cite{DBLP:conf/mm/AlbanieNVZ18} apply transfer learning to obtain supervision from another labeled modality. However, the improvement is very marginal. Data augmentation and semi-supervised learning through generative adversarial network \cite{GANS, Salimans2016Improved, Chang2017Learning} have also been explored. However, such models are hard to optimize due to the unstable training procedure and non-intuitive synthetic samples.

Inspired by the research in cross-modal retrieval task \cite{cao2017collective, zhang2016collaborative, wang2019semi}, in this paper, we propose a novel semi-supervised training strategy for discrete multi-modal emotion recognition.  We assume that different modalities are expected to express similar emotion information at the coarse-grained level (such as the utterance level) under a certain scenario as shown in Figure ~\ref{concept}. 
Under this assumption, we can regard this latent relationship as an auxiliary task to obtain guidance from unlabeled data to enhance the fully-supervised training procedure. Specifically, we use auto-encoder structure \cite{DAE} to extract utterance-level representation from different modalities and apply Maximum Mean Discrepancy(MMD) \cite{MMD} to restrict their distribution difference. 
We conduct extensive experiments to compare with other state-of-the-art techniques on two benchmark datasets, IEMOCAP \cite{IEMOCAP} and MELD \cite{poria2018meld},  and demonstrate the effectiveness of our proposed semi-supervised learning approach. We also carry out detailed analysis experiments to study the performance impact from each model component. 

The remainder of this paper is organized as follows. Section ~\ref{sec:related} introduces some related works. Section ~\ref{sec:model} describes the details of our emotion recognition system based on the proposed semi-supervised learning strategy, including the representation learning with DAE and the design of multiple loss functions. Section ~\ref{sec:expsetting} then shows the extensive experiment results on two benchmark datasets. Finally, Section ~\ref{sec:conclusion} presents our conclusions. 


\section{Related Works}
\label{sec:related}

\begin{figure*}[ht]
    \setlength{\abovecaptionskip}{4pt}
    \centering
    \includegraphics[width=1.0\linewidth]{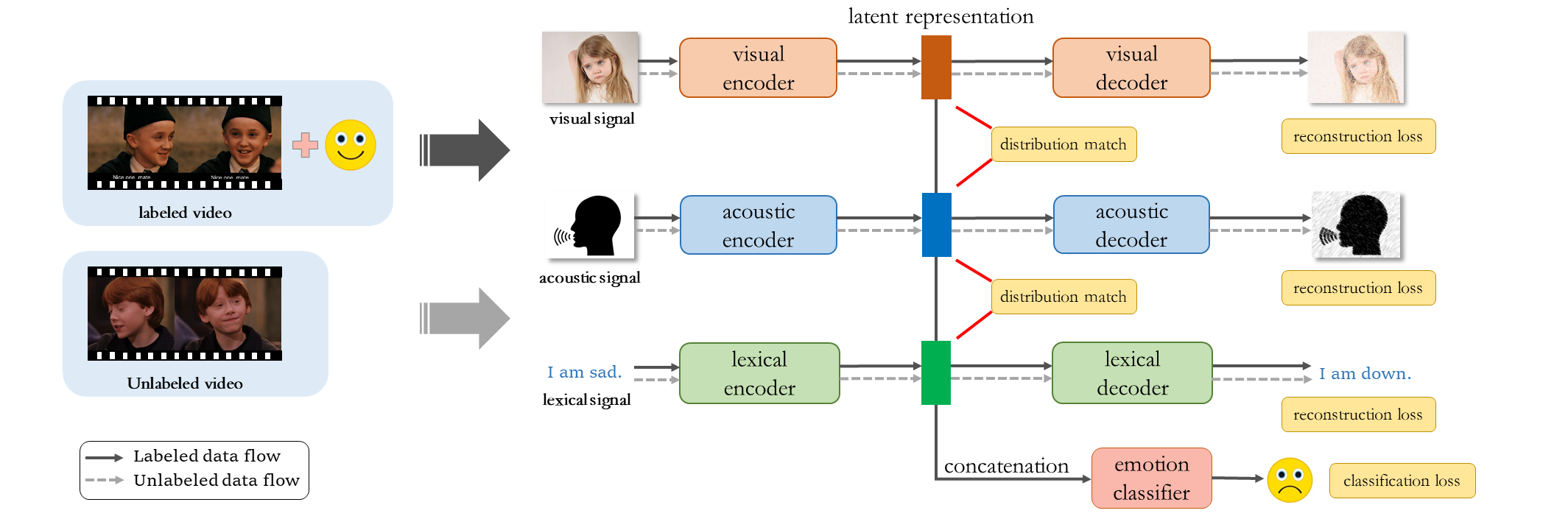}
    \caption{The overall structure of the proposed semi-supervised multi-modal emotion recognition framework. Both labeled and unlabeled data participate in the model learning (solid line for labeled data, dash line for unlabeled data). }
    \label{joint-network}
    \vspace{8pt}
\end{figure*}

\subsection{Multi-modal Emotion Recognition}
The quality of multi-modal features plays a decisive role in emotion recognition. Thus previous works have explored effective features in acoustic, visual and lexical modalities for emotion recognition tasks. 
Brady et al. \cite{Brady2016Multi} derive high-level acoustic, visual and physiological features from the low-level descriptors using sparse coding and deep learning. Seng et al. \cite{seng2016combined} uses a mixture of rule-based and machine learning techniques upon prosodic and spectral features to determine the emotion state contained in the audio and visual signal. The granularity of these features varies from frame level to sentence level.

For modality aggregation, Viktor et al. \cite{Emotion} use early fusion to concatenate multi-modal features as the input for the inference models. But it ignores the mismatch between different modalities. Considering the inner relationship alignment, Yoon et al. \cite{yoon}  propose deep dual recurrent encoder to combine text information and speech signals concurrently to gain a better understanding of emotion recognition. Xu et al. \cite{didi} propose to learn the frame-level alignment between speech and text signal via attention mechanism. They both learn the similarity between these two modalities to compress acoustic sequence and align the speech with text. However, the length of speech sequence is much larger than text sequence so that the accurate alignment is quite hard to learn. To avoid this problem, we tend to use utterance acoustic feature in this work.

For modeling the long-term dependency in emotion expression, Mao et al.\cite{mao2019deep} aggregate the segment-level decisions to improve utterance-level classification. Li et al. \cite{li2019towards} propose an novel representation learning component with residual convolutional network, multi-head self-attention and global context-aware attention LSTM. Following their suggestion, we utilize self-attention mechanism to capture temporal information as well.

Besides, emotion recognition in conversational scenario has become a popular sub-task recently. It emphasizes the interaction contextual information extraction and modeling in a human dialogue. Several approaches have been proposed to capture contextual and speaker cues to assist emotion recognition \cite{CMN,ICON,BCLSTM,dialoguernn,zhang2019modeling}. Although our proposed work is applied in non-interactive scenario, we compare the emotion recognition performance on the same dataset with these interactive models as well. 

\vspace{-6pt}
\subsection{Semi-supervised Emotion Recognition} 
Du et al. \cite{du2018semi} propose a semi-supervised multi-modal generative framework with non-uniformly weighted Gaussian mixture posterior approximation for the shared latent variable. They use a conditional probabilistic distribution for the unknown labels in the semi-supervised classification algorithm.
Salimans et al.\cite{Salimans2016Improved} use generative adversarial networks to implement semi-supervised learning. And Chang et al.\cite{Chang2017Learning} apply similar GAN-based semi-supervised framework on acoustic representations learning and it helps to improve emotion recognition. 
Besides generative models, Albanie et al.\cite{albanie2018emotion} explore to transfer emotion label from one modality to the other modality assuming that the supervised annotation does exist in one modality.

\vspace{-6pt}
\subsection{Distribution Matching}
Distribution matching \cite{jegou2010product, gong2012iterative} has been proposed and developed for cross-modal retrieval recently. Several methods \cite{cao2017collective, zhang2016collaborative} are proposed to map distribution from different domains into a shared space so that the representations of similar distribution from different domains can be aligned. They use various similarity metrics such as Asymmetric Quantizer Distance (AQD) \cite{AQD} and Maximum Mean Discrepancy (MMD) \cite{MMD}. 
Wang et al.\cite{wang2019semi} use distribution matching loss to incorporate the labeled and unlabeled data into one framework simultaneously and apply the semi-supervised cross-modal retrieval. Although all these previous works address the cross-modal search problem, we find that this concept can be seamlessly integrated with the semi-supervised emotion recognition. So we conduct the experiment and prove the efficiency of distribution matching across modality.

\section{Method}
\label{sec:model}


We assume that a video database naturally consists of information from three modalities (acoustic, visual and lexical). Given a labeled video database $\{X^L,Y\}=\{(x^{a}_{i},x^{v}_{i},x^{l}_{i},y_{i})\}_{i=1}^{n^{L}}$ and an unlabeled video database $\tilde{X}^{uL}=\{(\tilde{x}^{a}_{i},\tilde{x}^{v}_{i},\tilde{x}^{l}_{i})\}_{i=1}^{n^{uL}}$, where
$x^a,x^v,x^l$ denote the feature representation from the acoustic, visual and lexical modalities respectively, $L$ and $uL$ are used to distinguish labeled and unlabeled data, and $n^{L}$ and $n^{uL}$ denote the size of the labeled and unlabeled dataset respectively, our goal is to involve the unlabeled data in training to improve model performance. 

Previous study \cite{IEMOCAP} has shown that the emotional status is kept unchanged during an utterance and the average duration of utterances in the dataset is 4.5 seconds. 
Rigoulot et al \cite{srivastava2015unsupervised} has also studied the time course for human emotion expression and found that 4 seconds of speech emotions can usually be classified correctly.
Based on such discovery, we make the following assumption: although the emotion expression is not necessarily aligned at the frame level across modalities, the overall emotional state should be similar at the coarse-grained utterance level. 
We can utilize this assumption to extract supervision from unlabeled data. During training, we improve the accuracy of classification on labeled data, and reduce the difference of inter modality distribution on both labeled and unlabeled data simultaneously. 

\begin{equation}
\begin{aligned}
 Objective &= Classification(X^L,Y)  \\
    &+ Reconstruction(X^L,\tilde{X}^{uL})  \\
    &+ Matching(X^L,\tilde{X}^{uL}) 
  \label{overall}
\end{aligned}
\end{equation}

As shown in Fig.~\ref{joint-network}, two types of data (labeled and unlabeled data) both participate in the model learning (solid line for labeled data, dash line for unlabeled data). The additional unlabeled data can help learn more robust and emotion-salient latent representation. We use Maximum Mean Discrepancy (MMD) \cite{MMD} to measure the distribution similarity, which is motivated by its previous success in transfer learning and feature representation learning. 

Formally, the training objective of our semi-supervised model (Eq.~\ref{overall}) consists of three components corresponding to the emotion classification, data reconstruction and data distribution matching respectively, among which only the emotion classification component requires labeled data while the other two components are unsupervised. We present the details of model architecture (Section ~\ref{sec:architecture}) and loss function design (Section ~\ref{sec:loss}) in following subsections. 

\subsection{Multi-modal Network Architecture}
\label{sec:architecture}
As the model structure is related to the feature characteristics
of each modality, we first introduce the features and then present the network design.

\vspace{6pt}
\noindent\textsf{\textbf{Multi-modal Features}}

\noindent We first extract raw features from acoustic, visual and lexical modalities respectively.

\begin{itemize}[leftmargin=*]
\item Acoustic:
We utilize the toolkit OpenSMILE \cite{opensmile} to extract the utterance-level features with the configuration of INTERSPEECH2010 \cite{IS10}.
The extracted feature vector is in 1,582 dimension.

\item Visual:
We utilize the state-of-the-art Dense Convolutional Neural Networks (DenseNet) \cite{DenseNet}  to extract the facial features. The DenseNet is pretrained on the FERPlus \cite{BarsoumICMI2016} dataset for facial expression recognition. We extract the 342 dimensional activation from the last pooling layer for each face image as in \cite{Chen2017}.

\item Lexical:
We use the state-of-the-art word embedding trained by Bidirectional Encoder Representation from Transformers  (BERT) \cite{bert}. Each word is represented as a 1,024-dimensional vector. 
\end{itemize}
We apply z-normalization on each dimension of the raw features to reduce data discrepancy.

\vspace{6pt}
\noindent\textsf{\textbf{DAE for Representation Learning}}

\noindent Deep Auto-encoder (DAE) is proposed to learn high quality latent representation by encoding and reconstructing its input data. It can capture the data manifolds smoothly without losing too much original information \cite{DAE}. Cross modal distribution matching is applied to avoid the latent representation collapsing into zero space.

\begin{figure} 
\centering
\includegraphics[width=.4\textwidth]{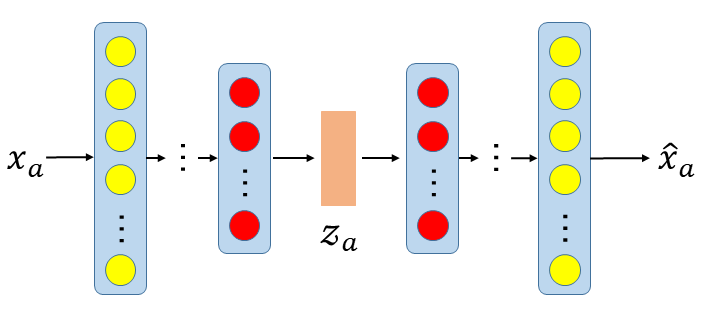}
\caption{Acoustic DAE: symmetric stacked linear layers.} 
\label{A_DAE} 
\end{figure}

\vspace{4pt}
\noindent\textbf{\textit{Acoustic:}} As the extracted acoustic features are at the utterance level, we consider the stacked linear layers as the encoder structure. We first transform the input acoustic feature $x_{a}$ to the latent representation $z_{a}$ with a set of linear layers and then get the reconstructed output $\hat{x}_{a}$ with symmetric layers. The network structure is shown in Figure \ref{A_DAE}. Please note that we do not apply frame-level distribution matching across modalities due to two reasons:
Firstly, as mentioned above, previous research has shown that emotion expression is not necessarily aligned at the frame level or word level across modalities  \cite{tsai2019multimodal, poria2017context, pham2019found, gu2018multimodal}, forcing the matching at the frame level will lead to inherently poor optimization of the model. Secondly, reconstruction at the frame level will result in very large amount of trainable parameters.


\begin{figure} 
\centering
\includegraphics[width=.4\textwidth]{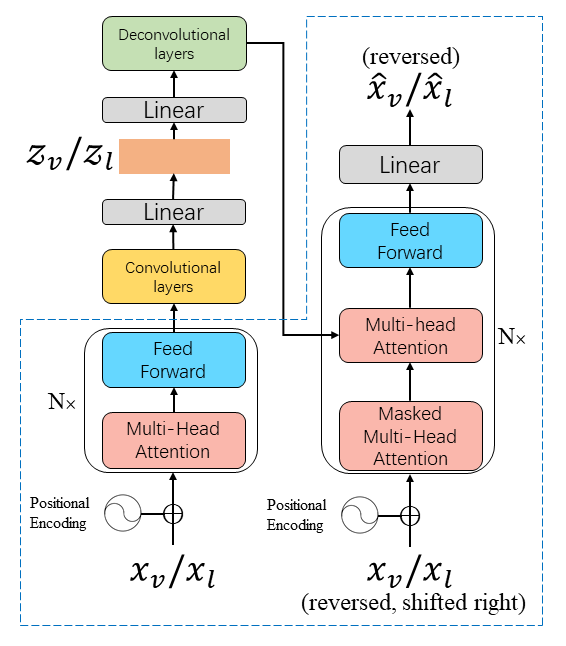}
\caption{Visual and Lexical DAE: modified Transformer structure. The part in blue dotted box is the original component in Transformer.} 
\label{L_DAE} 
\end{figure}

\vspace{4pt}
\noindent\textbf{\textit{Visual and Lexical:}} As both the raw visual and lexical features are a sequence of features, we consider seq2seq type of structure for the encoder and decoder. 
Transformer \cite{transformer} is one type of the state-of-the-art Seq2Seq models, which is completely based on attention mechanism and does not need recurrence and convolution. It captures the relative dependencies between elements of the sequence. We therefore design the transformer architecture for DAE of the visual and lexical modalities. 
The detailed component structure of visual and lexical DAE is shown in Figure \ref{L_DAE}. 

\begin{enumerate}[leftmargin=*]
\item Firstly, the input and output of the visual/lexical DAE model are the raw and reconstructed visual/lexical features, we therefore drop the embedding layer of the input and set the number of hidden units as 1 in the last linear output layer. 
\item Secondly, following Srivastava et al.\cite{srivastava2015unsupervised}, we take the reversed input sequence as the reconstruction target instead of the original sequence. Reversing the reconstruction target makes the optimization easier because the model can get off the ground by looking at low range correlations.
\item Lastly, the latent representations for different modalities are expected to follow the same shape. But the encoder output in the visual/lexical DAE is still stacked by time order. So in the middle of transformer, we set up a set of convolutional layers for down sampling and, symmetrically, a set of deconvolutional layers for reconstruction.
\end{enumerate}

\subsection{Loss Function Design}
\label{sec:loss}

The training objective of our semi-supervised model (Eq.~\ref{overall}) consists of three components corresponding to the emotion classification, data reconstruction and data distribution matching respectively.

\vspace{4pt}
\noindent\textbf{\textit{Reconstruction loss.}}
The loss function of reconstruction is Mean Squared Error (MSE) as follows:

\begin{equation}
\begin{aligned}
  L_{DAE}(x) = (x-rec(x))^2 
  \label{reconstruction_loss}
\end{aligned}
\end{equation}
\vspace{-10pt}

\noindent where $x$ is the input to DAE and $rec(x)$ is the reconstructed output from the corresponding DAE.

\vspace{4pt}
\noindent \textbf{\textit{Unsupervised Distribution Matching Loss.}}
Given a numbers of video samples segmented by utterance, we assume that the latent representation of acoustic, visual and lexical modalities from the same video can be mapped into a similar space, while the distribution of modalities from videos with different emotion status should be diverse. We employ Maximum Mean Discrepancy (MMD) \cite{MMD} to measure the distribution similarity. The distribution matching loss is as follows:

\vspace{-10pt}
\begin{equation}
\begin{aligned}
  L_{MMD}(p,q) &= \frac{1}{m(m-1)}\sum_{i\not=j}^m k(p_{i},p_{j}) \\ 
  &+ \frac{1}{n(n-1)}\sum_{i\not=j}^n k(q_{i},q_{j})
          - \frac{2}{mn}\sum_{i,j=1}^{m,n} k(p_{i},q_{j}) 
\label{mmd_loss}
\end{aligned}
\end{equation}

\begin{equation}
\begin{aligned}
  k(x,x^{'}) = exp(- \frac{ || x- x^{'}||^2}{2\sigma^2})
  \label{gaussion}
\end{aligned}
\end{equation}

\noindent where $p$, $q$ are the latent representation from two different modalities. The latent representation of each modality is mapped to Reproducing Kernel Hilbert Space (RKHS) before computing the distance. We use the Gaussian kernel (Eq.~\ref{gaussion}) to calculate the dot product in RKHS. This formula dose not contain any annotation information, so it can be forced on both labeled and unlabeled data to build up unsupervised training target.

\vspace{4pt}
\noindent \textbf{\textit{Supervised Emotion Classification Loss.}}
As we assume the emotion status is aligned across modalities at the utterance level, we can apply multi-modal fusion through directly concatenating the latent representation  and then feed it into the classifier. For the labeled data, we compute the cross-entropy loss (Eq.~\ref{classification}) for optimization.

\vspace{-10pt}
\begin{equation}
\begin{aligned}
  &L_{cls} = -\frac{1}{n^L}\sum_{i=1}^{n^L} \sum_{k=1}^{K} y_{i,k} log (p_{i,k}) \\ 
  &(p_{i,0},p_{i,1},... ,p_{i,K}) = \mathrm{softmax}(C([z^{a}_i;z^{v}_i;z^{l}_i]))
  \label{classification}
\end{aligned}
\end{equation}

\noindent where $C$ is a neural network based emotion classifier, $K$ is the number of emotion classes, $n^L$ is the total number of supervised samples, $y_i$ and $p_i$ are the annotated and predicted emotion class probability for input data $x_i$ and $[z^{a}_i;z^{v}_i;z^{l}_i]$ is the concatenation of the latent representation of $x_i$ from all the modalities.

\vspace{4pt}
\noindent \textbf{\textit{Joint Loss Function.}}
We combine all the losses into the joint loss function as below. For the supervised part, the loss function is set as:


\begin{equation}
    \begin{aligned}
        & L^{s} = L_{cls} + \alpha L_{rec}  + \beta L_{pair} \\  
        & L_{rec} =  L_{DAE}(z^a) + L_{DAE}(z^v) + L_{DAE}(z^l)         \\
        & L_{pair} = L_{MMD}(z^a,z^v) + L_{MMD}(z^a,z^l) 
    \end{aligned}
\end{equation}

However, we find that the latent representation still collapses into zero space. To avoid meaningless matching of modality representation, we add unpaired samples into the training. The `unpaired' means the feature extracted from different modalities doesn't belong to the same video or are not aligned. The gap between latent distribution of paired and unpaired should be obviously large. We shuffle the features across the acoustic, visual and lexical modalities so that they are not aligned anymore. We thus can form unpaired samples in this way. We hope the unpaired samples are mapped into different emotion space which means we enlarge the distribution distance during training. Based on such idea, the loss function is modified as:
\begin{equation}
\begin{aligned}
  &L^{s} = L_{cls} + \alpha L_{rec} + \beta(L_{pair} + L_{unpaired}) \\
  &L_{unpair} = - (L_{MMD}(z^a,z^v) + L_{MMD}(z^a,z^l)) 
\end{aligned}
\end{equation}
We then add unsupervised data in the training and the loss function for the unsupervised part is set as:
\begin{equation}
\begin{aligned}
  &L^{u} = \alpha L^{u}_{rec} + \beta(L^{u}_{pair} + L^{u}_{unpaired}) 
  \end{aligned}
\end{equation}
where $\alpha$, $\beta$ are hyper-parameters.
Finally, we form the semi-supervised loss function by combining the supervised and unsupervised losses:
\begin{equation}
\begin{aligned}
  &L^{semi} = L^{s} + \omega L^{u}
\end{aligned}
\end{equation}
where $\omega$ is the hyper-parameter to balance the two losses. 
\section{Experiment}
\label{sec:expsetting}

In this section, we present a series of comparison experiments on discrete emotion recognition task under fully-supervised and semi-supervised settings. 


\subsection{Data Description}
We utilize both labeled and unlabeled data for our experiments. 

\vspace{4pt}
\noindent\textbf{IEMOCAP}\cite{IEMOCAP} contains 12 hours of video recordings of situational dialogues. The videos are divided into five sessions. Each session contains only two actors so that in total there are ten actors in the database. The recorded dialogues are manually segmented into 10039 utterances with 9 discrete emotion classes, namely happiness, anger, sadness, fear, surprise, excitement, frustration, neutral and others. To compare with the state-of-the-art approaches, we follow the data split setting as in \cite{li2019towards} and use 5531 utterances from the top 4 emotion classes: happiness, anger, sadness and neutral (the `excitement' utterances are merged into the `happiness'  class). 
 The data  distribution is shown at Table~\ref{IEMOCAP_data}
We follow the speaker-independent setting to avoid actor overlap in the validation and testing set. Under this consideration, four sessions are chosen as the the training set and the remaining one session is divided into validation set and testing set. 

\begin{table}
\caption{Data distribution of IEMOCAP dataset}
\begin{center}
\begin{tabular}{c c c c c}
\toprule
 Happy &Anger  &Sadness  &Neutral &Total\\
\midrule
1636   &1103  &1084 &1708  &5531\\

\bottomrule
\end{tabular}
\label{IEMOCAP_data}
\end{center}
\end{table}

\vspace{4pt}
\noindent\textbf{MELD}\cite{poria2018meld} is a multi-modal conversational dataset. It extracts more than 1300 dialogues and 13000 utterances from Friends TV series with total 304 speakers. Each utterance segment contains audio track, visual scene and text transcript. And it is labeled with one of seven discrete emotions which are joy, anger, sadness, surprise, disgust, fear and neutral. The data  distribution is shown at Table~\ref{MELD_data}. 
The video contains multiple faces in a scene and the speaker label is not provided in the dataset. Thus, in this work, we can not match the speaker with his/her face exactly and we only use acoustic and lexical modalities in related experiments.

\vspace{4pt}
\noindent\textbf{AMI}\cite{AMI} dataset consists of about 100 hours of unlabeled meeting recordings. It provides video recordings of each speaker, voice track and transcripts of their speech. But there is no emotion annotation in the dataset, so we use it as the unsupervised dataset.

\begin{table}
\caption{Data distribution of MELD dataset}
\begin{center}
\begin{tabular}{c c c c c}
\toprule
Emotion &Train &Dev &Test &Total \\
\midrule
Anger    & 1109  & 153 & 345  &1607\\
Disgust  & 271   & 22  & 68   &361\\
Fear     & 268   & 40  & 50   &358\\
Joy      & 1743  & 163 & 402  &2308\\
Neutral  & 4710  & 470 & 1256 &6436\\
Sadness  & 683   & 111 & 208  &1002\\
Surprise & 1205  & 150 & 281  &1636\\
\bottomrule
\end{tabular}
\label{MELD_data}
\end{center}
\end{table}

\subsection{Implementation Details }
\vspace{4pt}
\noindent\textbf{Unlabeled Sample Sampling:} 
We summarize several basic rules to select unsupervised data and perform semi-supervised learning: firstly, the spoken language, cultural background and age range of speakers should be similar. Secondly, the camera setting should be as consistent as possible (e.g. lighting, shooting position, resolution, etc.). To avoid the case that the sampled region from AMI is silence, we randomly select three continuous words in the transcript and look up the time region in the video. We then extract the audio and video segment with middle word as the center of the segment.
Due to various video duration and utterance length in AMI and IEMOCAP datasets, we pre-define the video duration of unlabeled sample in advance. Because the duration of 80\% utterances in the IEMOCAP dataset is limited to 7.2s, we therefore extract unlabeled sample with 7.2s as the crop width for the experiments on the IEMOCAP dataset. Similarly, for experiments on the MELD dataset whose utterances are shorter, the crop width is set as 3.5s. This step ensures that there is no significant difference in the duration between labeled samples and unlabeled samples.
Additionally, people tend to keep neutral emotion status during meetings and less likely to express sad or anger emotions. We apply a pre-trained vanilla emotion classifier with IEMOCAP dataset on the AMI dataset and observe that  about 80\% of the 20,000 unlabeled samples are classified as happiness and neutral. We therefore apply sub-sampling on these two emotion types which selects 5000 samples for the happiness and neutral classes respectively. The number of samples for sadness and anger classes is less than 5000, we therefore apply over-sampling on these two classes and collect 5000 samples for each class respectively. After the filtering process, the unlabeled dataset is more balanced.

\begin{table}
\caption{Model architecture setting. In convolutional and deconvolutional layers, we denote kernel size as $k$, stride length as $s$, padding length as $p$ and the number of channels as $c$. In Transformer, we denote the number of self-attention heads as $h$, the number of transformer blocks as $b$ and the size of hidden embedding as $e$.}

\begin{center}
\begin{tabular}{c|c|c}
    \toprule
    Modality   &Input   &Encoder  \\
    \hline
    A  &$1\times1582$ &Linear:$1582\rightarrow512\rightarrow256\rightarrow128$ \\
    \hline
    \multirow{7}{*}{V} &\multirow{6}{*}{$18\times342$}  & Transformer Encoder $h$=4,$b$=2,$e$=342\\
                   \cline{3-3}
                 & &Convolutional layers                       \\
                 & &$k$=4,$s$=2,$p$=1,$c$=16                 \\
                 & &$k$=5,$s$=2,$p$=1,$c$=64                 \\
                 & &$k$=3,$s$=3,$p$=1,$c$=32                 \\
                 \cline{3-3}
                 & &Flatten layer                    \\
                 & &Linear:$1856\rightarrow128$                  \\
    \hline
    \multirow{6}{*}{L} &\multirow{6}{*}{$22\times1024$}& Transformer Encoder $h$=4,$b$=2,$e$=1024\\
                                                     \cline{3-3}
                       &                             & convolutional layers \\
                       &                             & $k$=4,$s$=2,$p$=1,$c$=64     \\
                       &                             & $k$=4,$s$=3,$p$=1,$c$=4      \\
                                                     \cline{3-3}
                       &                             & Flatten layer        \\
                       &                             & Linear:$2736\rightarrow512\rightarrow128$  \\
                                                                             
    \hline

       &Input   &Decoder   \\
    \hline
    A  &$1\times128$  &Linear:$128\rightarrow256\rightarrow512\rightarrow1582$ \\
    \hline
    \multirow{7}{*}{V} &\multirow{6}{*}{$1\times128$}  & Linear:$128\rightarrow1856$ \\
                 &     & Reshape to $32\times2\times29$ \\
                       \cline{3-3}
                 &     & Deconvolutional layers \\
                 &     & $k$=3,$s$=3,$p$=1,$c$=64 \\

                 &     & $k$=5,$s$=2,$p$=1,$c$=16 \\
                 &     & $k$=4,$s$=2,$p$=1,$c$=1 \\
                 \cline{3-3}
                 &     &Transformer Decoder $h$=4,$b$=2,$e$=342  \\
    \hline
    \multirow{6}{*}{L} &\multirow{6}{*}{$1\times128$}  &Linear:$128\rightarrow512\rightarrow2736$\\
                       &        &Reshape to $4\times4\times171$ \\
                                \cline{3-3}
                       &        &Deconvolutional layers \\
                       &        &$k$=4,$s$=3,$p$=1,$c$=64 \\
                       &        &$k$=4,$s$=2,$p$=1,$c$=1 \\
                        \cline{3-3}
                       &         &Transformer Decoder $h$=4,$b$=2,$e$=1024\\
                                                                             
    \bottomrule
\end{tabular}
\label{network-hyperparameters}
\end{center}
\end{table}

\vspace{4pt}
\noindent \textbf{Face Extraction:} We apply face detection and extraction on all the datasets with the toolkit Seetaface \cite{seetaface}. Each face image is transformed into the gray scale with size of 64x64. These videos contain 30 frames per second and there is almost no change between adjacent frames. To reduce computation cost without losing too much information, we set the sampling rate to 1/10, which means we can get about 3 frames per second. For those frames where faces cannot be detected, we use the detection results of the previous frame.
However, for the AMI dataset, there are about 11\% videos where faces can not be detected. We simply drop these samples. Finally, we get 20,000 unlabeled samples from AMI in total.



\vspace{5pt}
\noindent\textbf{Hyper-parameters:} For the IEMOCAP dataset, the number of face images and words in one utterance is fixed as 18 and 22. For the MELD dataset, the number of words in  a single utterance is fixed as 12. We pad zeros when the utterance is not long enough and cut off if it is too long. We set the batch size as 128, the weight of reconstruction loss $\alpha$ as 0.2, the weight of MMD loss $\beta$ as 0.1 and the weight of unsupervised part $\omega$ as 0.3. We apply Adam algorithm with learning rate of 1e-3 to optimize the parameters. The detailed setting of network is presented in Table~ \ref{network-hyperparameters}. The structure of the encoder and decoder is completely symmetric. We select the best model based on 5-fold cross validation on the validation set and report its performance on the testing set.



\subsection{Comparison Experiments and Results}
\label{sec:expresults}
\vspace{-4pt}
We first compare the proposed semi-supervised framework with several recent state-of-the-art approaches on the IEMOCAP dataset. 

1): Yoon et al. \cite{yoon} propose a deep dual recurrent encoder to combine the text information and audio signals. They first investigate the performance of uni-modal recurrent encoder on audio and text (\textbf{ARE}, \textbf{TRE}). Then they propose multi-modal dual recurrent encoder with and without attention techique (\textbf{MDRE+Att}, \textbf{MDRE}). All the results are reported under speaker-independent settings.
    
2): Xu et al. \cite{didi} propose to learn the frame-level alignment between audio and text via the attention mechanism in order to produce more accurate multi-modal feature representations. They conduct uni-modal experiments on acoustic and lexical data using LSTM with attention (\textbf{LSTM+Att}). For multi-modal settings, they compare the performance of direct concatenation (\textbf{Concat}) and applying alignment (\textbf{Alignment}) via attention computing. 
    They report the results under speaker-independent settings as well.

For fair comparison with above baselines, we also conduct speaker-independent experiments and report the performance with weighted average precision (WAP) and unweighted accuracy (UA) \cite{kent1955machine}.

\begin{table}
    \caption{Fully-supervised experiment results under speaker-independent settings on IEMOCAP}
    
    \centering
    \begin{tabular}{c c c c}
    \toprule
    Modality & Model   &WAP &UA \\
    \hline
    \multirow{2}*{A} & ARE\cite{yoon} &54.6\% &58.0\% \\
    & LSTM+Att\cite{didi} &55.5\%  &57.4\% \\
    & \textbf{Ours} &\textbf{57.2}\% &\textbf{58.5}\% \\
    \hline 
    V & \textbf{Ours}  &\textbf{52.5\%}  &\textbf{45.4\%} \\
    \hline
    \multirow{3}*{L} &
    TRE\cite{yoon} &63.5\% &59.1\% \\
    & LSTM+Att\cite{didi} &59.0\% &57.8\% \\
    & \textbf{Ours} &\textbf{65.4\%}  &\textbf{64.8\%} \\ 
    \hline

    A+V &\textbf{Ours} &\textbf{63.2\%} &\textbf{60.8\%}   \\
    \hline 
    \multirow{6}*{A+L} &
    MDRE\cite{yoon}      & \textbf{71.8}\%    &67.2\%         \\
    & MDRE+Att\cite{yoon}   &69.0\%    &68.1\%         \\  
    & Concat\cite{didi}      &67.1\%    &67.7\%    \\
    & Alignment\cite{didi}   &68.4\%    & \textbf{70.9}\%    \\
    & \textbf{Ours}    &70.3\%     &68.6\%   \\
    \hline
    A+V+L &\textbf{Ours} &\textbf{73.0\%} &\textbf{72.5\%}   \\

    \bottomrule

    \end{tabular}
    \label{Speaker-independent-part1}
\end{table}

\begin{table}[t]
    \caption{Semi-supervised experiment results under speaker-independent settings on IEMOCAP}

    \centering
    \begin{tabular}{c c c c}
    \toprule
    Modality & Supervised mode   &WAP &UA \\
    \hline
    \multirow{2}*{A+V}
    & Ours-fully &63.2\% &60.8\%   \\
    & \textbf{Ours-semi}     &\textbf{63.7\%}   &\textbf{61.2\%}     \\
    \hline 
    \multirow{2}*{A+L}
    & Ours-fully         &70.3\%     &68.6\%   \\
    & \textbf{Ours-semi} &\textbf{72.6\%}    &\textbf{72.1\%}    \\
    \hline
    \multirow{2}*{A+V+L} &Ours-fully &73.0\% &72.5\%   \\
    & \textbf{Ours-semi}   &\textbf{75.6\%}  &\textbf{74.5\%}      \\
    \bottomrule

    \end{tabular}
    \label{Speaker-independent-part2}
\end{table}

\begin{figure}
\centering
\includegraphics[width=0.6\linewidth]{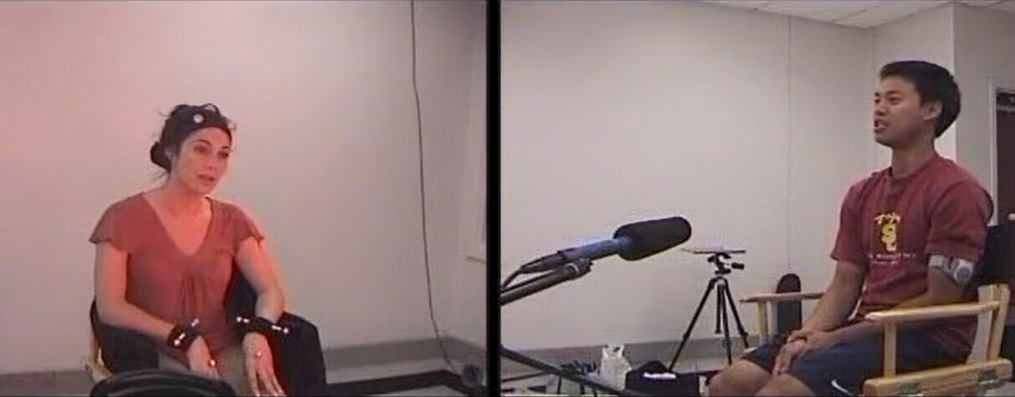}
\caption{An example frame from videos in IEMOCAP dataset. We can not capture the full face of the right person which reduces the capability of visual model.}
\label{side_face}
\end{figure}

Table~\ref{Speaker-independent-part1} presents the experiment results under the fully supervised speaker-independent setting. 
We can see from the results that our method outperforms recurrent encoder and LSTM with attention under the uni-modal scenario, which indicates that the feature selection and DAE architecture can capture emotional characteristics well and the long-term dependency in transcript can be modeled effectively. 
The visual modality achieves worse performance compared with acoustic and lexical modalities. The possible reason might be that nearly half of the speakers only show part of their faces as exampled by the speaker on the right in Fig~\ref{side_face}.
The low quality of face images limits visual model performance, which leads to that few research explores visual modality on the IEMOCAP dataset. The better performance of multi-modal model demonstrates that multiple modalities are complementary to each other for emotion expression.


The semi-supervised results are shown in Table~\ref{Speaker-independent-part2}. We compare the performance between fully-supervised setting and semi-supervised setting to verify the feasibility of our assumption of modality distribution matching. Since the proposed semi-supervised approach needs at least two modalities, we didn't implement semi-supervised experiments for uni-modal settings. 
Our semi-supervised training strategy boosts the model capability in all modality combination, which outperforms all the baseline approaches. It demonstrates the rationality and effectiveness of the proposed assumption and model, which can take advantages of unsupervised data and extract more emotion-salient latent representation. 

\begin{table}
	\caption{Performance comparison (weighted F1 score) on MELD dataset. * indicates that the corresponding approach uses conversation context information, and $\triangle$ indicates that the corresponding approach uses the speaker information}

	\begin{center}
		
		\begin{tabular}{c c c c}
			\toprule
			Approaches                            &A          &L        &A+L           \\
			\hline
			MFN\cite{MFN}                          &-          &-        &54.7\%        \\
			CMN\cite{CMN} $\triangle$               &38.3\%     &54.5\%   &55.5\%        \\
			ICON\cite{ICON} *$\triangle$          &37.7\%     &54.6\%   &56.3\%        \\
			BC-LSTM\cite{BCLSTM} *                  &36.4\%     &54.3\%   &56.8\%        \\
			DialogueRNN\cite{dialoguernn} *$\triangle$               &34.0\%     &55.1\%   &57.0\%        \\
			ConGCN\cite{zhang2019modeling} without context$\triangle$     &39.2\%     &54.3\%   &57.1\%        \\
			ConGCN\cite{zhang2019modeling} without speaker*               &37.3\%     &55.3\%   &57.4\%        \\
			ConGCN\cite{zhang2019modeling} *$\triangle$        &42.2\%     &57.4\%   &59.4\%        \\
			\hline
			Ours-fully      &40.2\%     &53.0\%   &56.1\%        \\
			Ours-semi       &-          &-        &57.1\%        \\
			
			\bottomrule
		\end{tabular}
		\label{meld}
	\end{center}
\end{table}

We then present the experiment results on the MELD dataset. Because MELD dataset is the new emotion dataset collected in the interaction scenario, majority of existing approaches validating on it consider auxiliary information from interaction, such as interaction context or speaker information. However, in this work, our focus is on how to effectively utilize unlabeled data, our proposed approach doesn't consider the context or speaker information. 
So in this set of experiments, we not only compare to approaches without using interaction related information, but also to those interactive approaches. We use following state-of-the-art baselines for comparison.

\begin{enumerate}[leftmargin=*]
    \item Zadeh et al. propose Memory Fusion Network (\textbf{MFN}) \cite{MFN} which focuses on improving multi-modal fusion effectiveness. This method does not use any context or speaker information.
    \item Poria et al. propose Bidirectional Contextual LSTM (\textbf{BC-LSTM}) \cite{BCLSTM}. It performs contextual information fusion in conversational scenario.
    \item Hazarika et al. propose Conversational Memory Network (\textbf{CMN}) \cite{CMN} and Interactive Conversational Memory Network (\textbf{ICON}) \cite{ICON}. Both models utilize the context of the speaker and the interlocutor during the two-speaker interaction. The former ignores global contextual information while the later incorporates the global context.
    \item Majumder et al. propose \textbf{DialogueRNN} \cite{dialoguernn} which models preceding emotion status of two speakers and global contextual information through three GRUs.
    \item Zhang et al. propose Context-sensitive and speaker-sensitive graph-based
    convolutional neural network (\textbf{ConGCN}) \cite{zhang2019modeling} to simulate dialogue relationship. They aggregate multi-speaker and multi-conversation into a graph and explore the latent connection. And they also report the results of experiment without conversation-sensitive component.
\end{enumerate}
Since the number of each emotion category is unbalanced in the MELD dataset, following Zhang et al. \cite{zhang2019modeling}, we report performance with weighted average F1 score \cite{maf1}.

As shown in Table~\ref{meld}, our semi-supervised model significantly outperforms MFN which also does not use any conversation context and speaker information. 
Our model also achieves better performance than CMN, ICON and BC-LSTM, which uses additional auxiliary information, either conversation context information or speaker information. It demonstrates the advantage of our model in isolated emotion recognition scenario. 
Furthermore, our model achieves comparable performance with DialogueRNN and ConGCN, the two state-of-the-art interactive emotion recognition approaches proposed recently. 
Although ConGCN outperforms our model by 2.3\% when it makes full usage of both speaker and contextual information, the overall experiment results show that even though our model does not utilize any auxiliary interaction information, it is also very competitive in conversational scenarios. 

\subsection{Analysis and Discussion}

\begin{figure}
\centering
\begin{subfigure}[]{
\label{fig:side:a}
\includegraphics[width=1.6in]{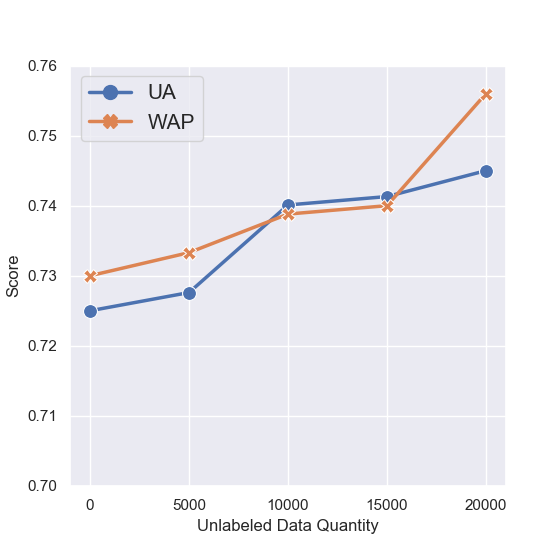}}
\end{subfigure}
\centering
\begin{subfigure}[]{
\label{fig:side:b}
\includegraphics[width=1.6in]{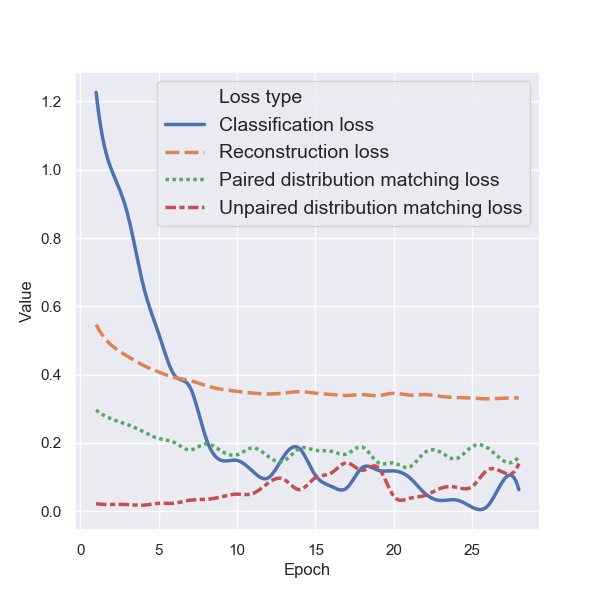}}
\end{subfigure}
\caption{(a) Influence of unlabeled data quantity. (b) The changing curves of loss value.}
\end{figure}

\textbf{Unlabeled data quantity analysis.} To gain more insights about the impact of unlabeled data, we analyze the classification performance change with different quantity of unlabeled training data. We conduct the semi-supervised experiment with combining all the three modalities. To show the impact of unlabeled data, we keep the hyper-parameter unchanged except the number of unlabeled data. We train a fully-supervised model at first and then add 5000 unlabeled samples step by step till all the 20,000 unlabeled samples from the AMI corpus are used. 
As shown in Fig.~\ref{fig:side:a}, the performance of semi-supervised model gradually improves with the increase of unlabeled samples, which indicates that the additional samples benefit the generalization and robustness of the recognition model. 

\begin{figure}
	\centering
	\begin{subfigure}[]{	
			\label{fig:cm:a}	
			\includegraphics[width=1.6in]{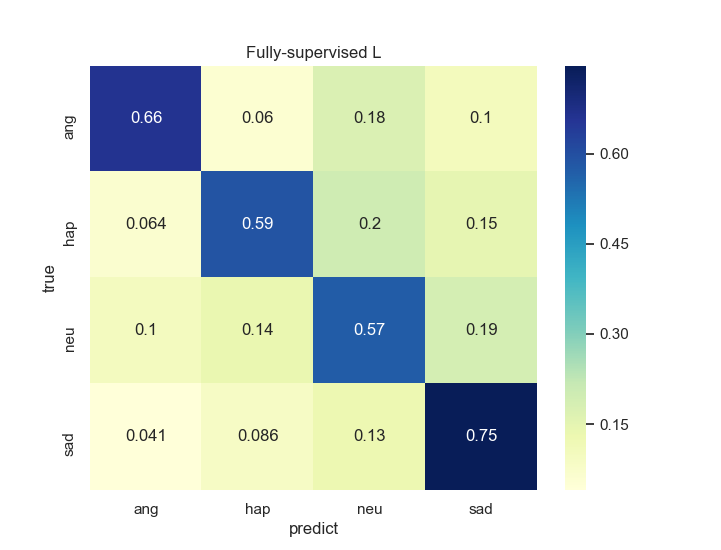}}
	\end{subfigure}%
	\begin{subfigure}[]{
			\label{fig:cm:b}
			\includegraphics[width=1.6in]{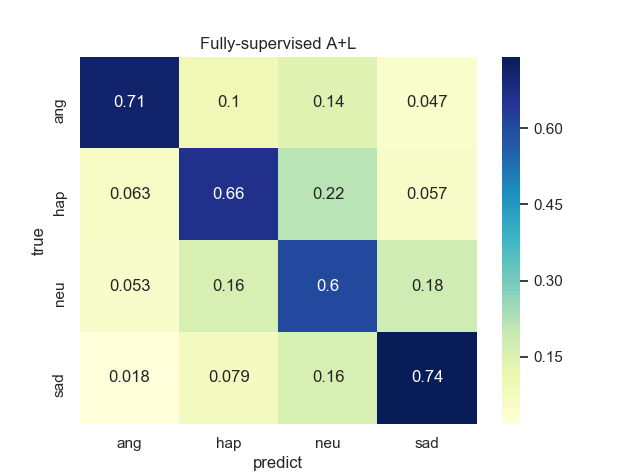}}
	\end{subfigure}
	\begin{subfigure}[]{	
			\label{fig:cm:c}	
			\includegraphics[width=1.6in]{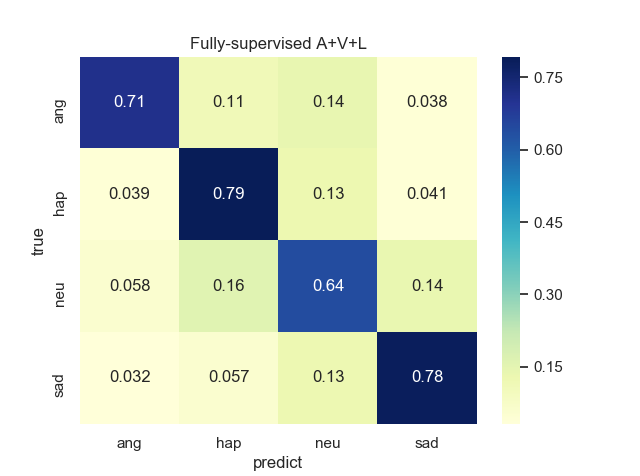}}
	\end{subfigure}%
	\begin{subfigure}[]{
			\label{fig:cm:d}
			\includegraphics[width=1.6in]{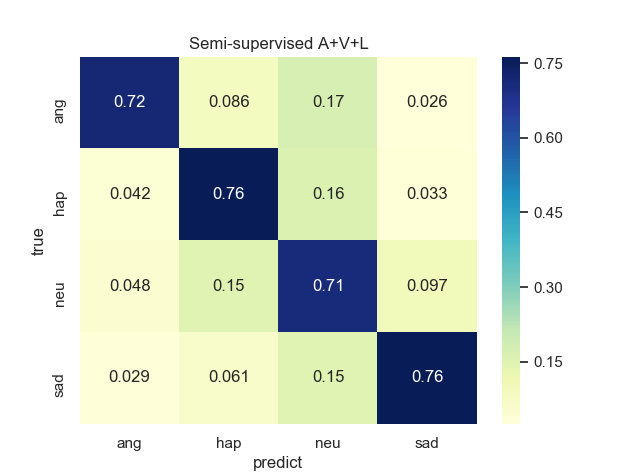}}
	\end{subfigure}
    
	\vspace{-4pt}
	\caption{Confusion matrix of experiments on IEMOCAP datasets. (a) experiments on lexical modality; (b) fully-supervised experiment on acoustic and lexical modalities; (c) fully-supervised experiments on acoustic,visual and lexical modalities; (d) semi-supervised experiments on acoustic, visual and lexical modalities}
	\label{cm}
\end{figure}

\textbf{Confusion matrix analysis.} As shown in Fig.~\ref{cm}, neutral samples are more likely to be identified as emotional categories. According to the confusion matrices, multi-modal combination boosts the performance on all classes and semi-supervised learning mainly improves performance on the neutral class.   

\textbf{Training procedure analysis.} We present the loss curves in Fig.~\ref{fig:side:b}. We can see that the classification loss, our main optimization target, decreases rapidly and converges within 15 epochs. Due to the limited scale of training data, most of the best models on the validation set appear between 10th and 15th epoch in our experiments. Too many training iterations will lead to over fitting on the training set. The curve of reconstruction loss changes smoothly and converges stably. The change of distribution loss also meets our expectation that the paired one and the unpaired one change towards the opposite direction and then converge to a similar value.

\section{Ablation Study}

To explore the contribution of each loss component, we fix the supervised emotion classification loss and do an ablation study on reconstruction loss and unsupervised distribution matching loss. We select the speaker-independent experiment of the acoustic, visual and lexical modalities on the IEMOCAP dataset as example. Table ~\ref{A+V+L} presents the results.  The experiments are divided into fully-supervised part and semi-supervised part. In the fully-supervised experiments, the model only employing distribution matching loss achieves worst performance. 
During the model structure design, we have a concern that distribution matching loss may harm the training with zero collapsing. Since the model aims to narrow the gap between two distributions, one trick is to drop most information and map all the distributions into a simple space (e.g. zero space). This experiment verifies our concern so that we need to avoid this misguidance. After we add the reconstruction target, the model improves and outperforms the vanilla model. When both reconstruction and distribution matching are employed, it further boosts the recognition performance. For the semi-supervised model, the distribution matching component is essential, so we only remove the reconstruction component. Similar to the fully-supervised experiment, the single distribution matching target misleads the training process and gets worse performance compared with the vanilla model. According to this study, distribution matching needs to be constrained to provide proper supervision for training. 
 
\begin{table}

\caption{Experimental results for component contribution evaluation (based on A+V+L)}
\begin{center}

\begin{tabular}{c c c | c c}
    \toprule
    Setting       &Reconstruction  &MMD   &WAP &UA \\
    \hline
    \multirow{3}{*}{fully-supervised}    &\ding{55}   &\ding{55}   &72.7\%  &71.5\% \\  
                                        &\ding{55}   &\ding{51}  &72.0\%  &71.5\% \\  
                                        &\ding{51}  &\ding{55}   &72.5\%  &71.7\% \\  
                                        &\ding{51}  &\ding{51}  &73.0\%  &72.5\% \\  
    \hline
    \multirow{2}{*}{semi-supervised} &\ding{55}  &\ding{51}   &72.9\%  &71.5\% \\  
                                     &\ding{51}  &\ding{51}   &75.6\%  &74.5\% \\
    \bottomrule
\end{tabular}
\label{A+V+L}
\end{center}
\end{table}

\section{Conclusion}
\label{sec:conclusion}

In this work, under the assumption that the emotion status is consistent across different modalities at the coarse utterance level, we propose a novel semi-supervised learning method based on cross-modal distribution matching for multi-modal emotion recognition. We jointly optimize the emotion classification, utterance-level cross-modal distribution matching and feature reconstruction objectives. Extensive experiments on IEMOCAP and MELD datasets prove the effectiveness of our proposed semi-supervised model and demonstrate that unlabeled data and multi-modality fusion both benefit the classification performance. Our model without contextual information outperforms existing state-of-the-art models in non-interactive scenario and is competitive with interactive methods. 

\section{Acknowledgement}
This work was supported by National Natural Science Foundation of China (No. 61772535) and Beijing Natural Science Foundation (No. 4192028).

\bibliographystyle{ACM-Reference-Format}
\bibliography{sample-base}

\end{document}